\documentclass[aps,prl,amsmath,amssymb,preprint,superscriptaddress]{revtex4-1}
\usepackage{graphicx}

\begin{document}
\title{Microwave assisted coherent and nonlinear control in cavity piezo-optomechanical system}

\author{King Yan Fong}
%\email{kingyan.fong@yale.edu}
\affiliation{Department of Electrical Engineering, Yale University, New Haven, CT 06511, USA}
\author{Linran Fan}
\affiliation{Department of Electrical Engineering, Yale University, New Haven, CT 06511, USA}
\author{Liang Jiang}
\affiliation{Department of Applied Physics, Yale University, New Haven, CT 06511, USA}
\author{Xu Han}
\affiliation{Department of Electrical Engineering, Yale University, New Haven, CT 06511, USA}
\author{Hong X. Tang}
\email{hong.tang@yale.edu}
\affiliation{Department of Electrical Engineering, Yale University, New Haven, CT 06511, USA}

\date{\today}

\begin{abstract}
We present a cavity piezo-optomechanical system where microwave and optical degrees of freedom are coupled through an ultra-high frequency mechanical resonator. By utilizing the coherence among the three interacting modes, we demonstrate optical amplification, coherent absorption and a more general asymmetric Fano resonance. The strong piezoelectric drive further allows access to the large-amplitude-induced optomechanical nonlinearity, with which  optical transparency at higher harmonics through multi-phonon scattering is demonstrated. 
\end{abstract}

\pacs{07.10.Cm, 42.50.Wk, 42.82.Fv, 85.85.+j}

\maketitle

Cavity optomechanics is the study of the system where an optical cavity is coupled with a mechanical resonator. It has been an active research topic over the past few decades while many breakthroughs in experiment have emerged in recent years. Interesting classical and quantum phenomena such as regenerative mechanical oscillation \cite{Vahala_OE_13_2005, Vahala_PRL_95_2005}, chaotic dynamics \cite{Vahala_PRL_98_2007}, optomechanically induced transparency (OMIT) \cite{Kippenberg_Science_2010_OMIT, Simmonds_Nature_2011_StrongCoupling, Painter_Nature_2011_OMIT, Vitali_PRA_2013_OMIT}, sideband cooling to quantum mechanical ground state \cite{Simmonds_Nature_2011_GroundState, Painter_Nature_2011_GroundState}, and squeezing of light below shot noise limit \cite{StamperKurn_Nature_2012_Squeezing, Painter_Nature_2013_Squeezing, Regal_PRX_2013_Squeezing}, have been observed experimentally. A comprehensive review of the field can be found in Ref. \cite{Florian_Arxiv_2013_OMReview}. From a more general perspective, the optical mode used in the system is not limited to optical photon but can also be electromagnetic resonance at relatively low frequency such as the microwave mode in superconducting cavities. Recently, there has been a rising interest of combining both optomechanics and electromechanics to realize hybrid opto-electro-mechanical systems where optical and RF/microwave cavities are coupled through a common mechanical resonator \cite{Lehnert_JP_2011, Bowen_PRL_2010, Cleland_NatPhys_2013, Tang_APL_2012_SlotPhotonicCrystal, Tang_APL_2013_Piezooptomechanics, Tang_APL_2013_OMCrystal, Bhave_Transducers_2013, Bhave_NL_2013, Polzik_Nature_2014_RFOpticsConversion}. Such a hybrid system finds applications in microwave photonics, high frequency oscillators \cite{Bhave_Transducers_2013} and promises to realize microwave-optical photons interconversions \cite{Lehnert_JP_2011, Bhave_NL_2013, Polzik_Nature_2014_RFOpticsConversion}, phonon-mediated electromagnetically induced absorption \cite{Agarwal_PRA_2013_EIAinOEM}, and entanglement between optical and microwave photons \cite{Milburn_PRA_2011}.

In this letter, we develop a cavity piezo-optomechanical system using aluminum nitride (AlN) micro-disk resonator, where the optical and microwave modes are coupled to the mechanical mode through radiation pressure and piezoelectric force. By making use of the coherence among the three interacting modes, we demonstrate coherent absorption and amplification, and a more general asymmetric Fano resonance which can be tuned through the whole phase-space. With the strong piezoelectric actuation, we are able to observe the large-amplitude-induced nonlinear optomechanical response, which allows us to further demonstrate high-harmonics optical transparency through multi-phonon scattering.

\begin{figure}[t]
\centering
\includegraphics[width=12cm]{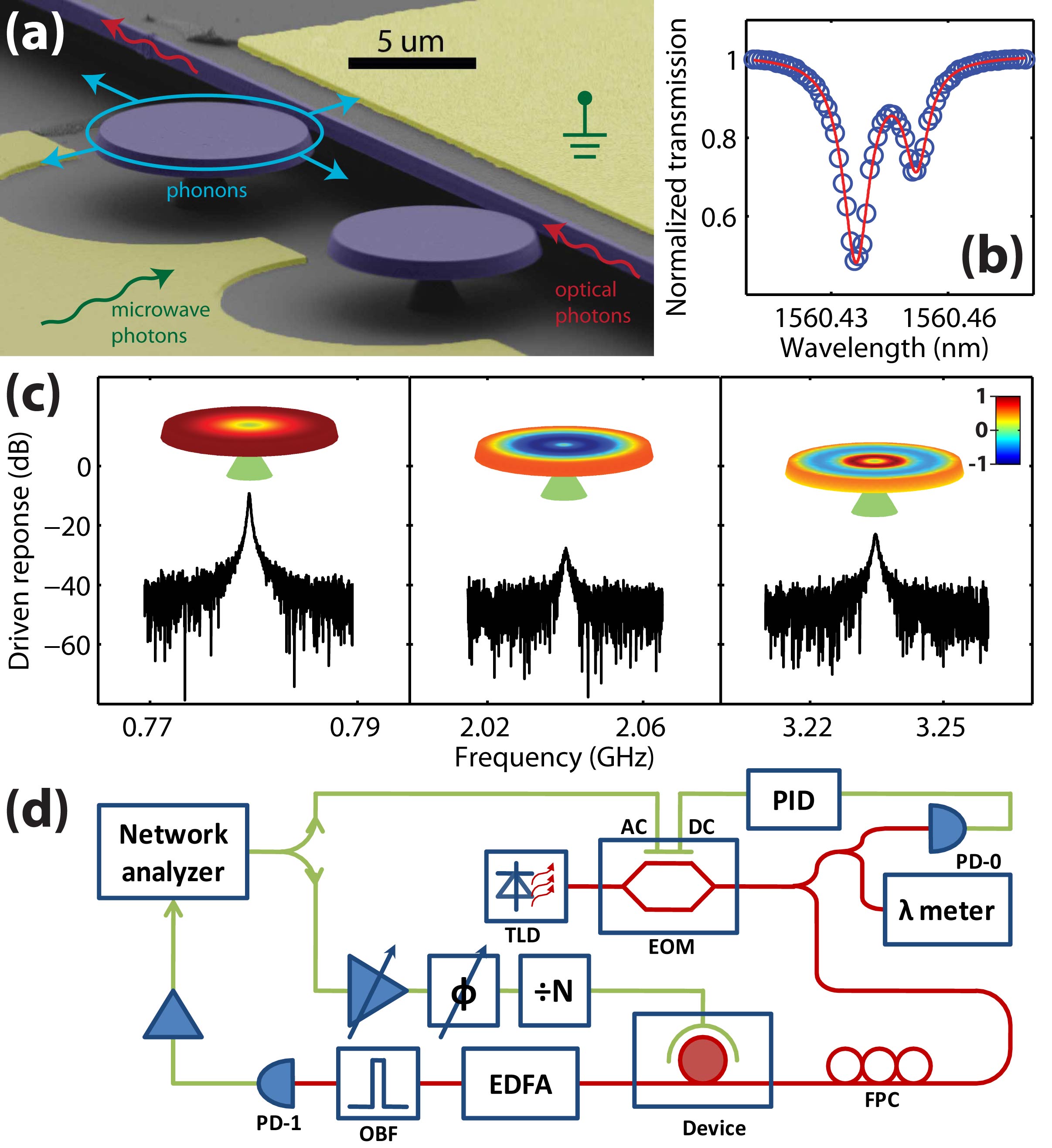}
\caption{(a) SEM image of the device showing the micro-disk resonator and the coupling waveguide (purple), and the gold electrodes (yellow). (b) Normalized optical transmission (c) Power spectra of the first three mechanical radial-contour modes. Insets show the finite element simulation of the normalized radial displacement profiles. (d) Schematics of the measurement setup. }
\label{fig:device_setup}
\end{figure}

Fig. \ref{fig:device_setup} (a) shows the SEM image of the device under study. The device consists of a suspended micro-disk resonator which supports optical and mechanical modes, and an integrated electrode connected to a transmission line which carries microwave mode. The micro-disk resonator made of AlN has a radius of 4.4 $\mathrm{\mu m}$ and a thickness of 650 nm. A coupling waveguide is fabricated next to the micro-disk for optical input and output. The micro-disk is sandwiched between two integrated electrodes which are connected to a microwave transmission line. The microwave field penetrates into the micro-disk and actuate the mechanical mode piezoelectrically.

Fig. \ref{fig:device_setup} (b) shows the transmission spectrum of an optical resonance, which has a double-dip feature due to mode-splitting of the originally degenerate clockwise and counter-clockwise modes \cite{Haroche_OL_1995}. The resonance on the shorter (longer) wavelength side has a cavity dissipation rate of $\kappa/2\pi=1.02$ GHz (0.931 GHz). At room temperature and under atmospheric pressure, we characterize the mechanical resonances by actuating the device piezoelectrically. The measured power spectra are plotted in Fig. \ref{fig:device_setup} (c). The first three mechanical radial-contour modes are observed at 779.6 MHz, 2.040 GHz, and 3.235 GHz, which agrees well with the results of finite element simulation. The simulated normalized radial displacement profiles of the three modes are shown in the corresponding insets. The phonon dissipation rates of the three modes are measured as $\gamma/2\pi=202$ kHz, 1.25 MHz, and 961 kHz. The measured displacement shift due to piezoelectric actuation is $dR/dV_{dc}=4.5$ fm/V, which agrees with numerical simulation result. The actuation efficiency is relatively small compared to other demonstrated system \cite{Cleland_NatPhys_2013} because of the larger electrode separation used.

The piezoelectric effect in AlN naturally couples the microwave mode to the mechanical mode by introducing an eletro-mechanical coupling energy given by $H_{em}=\int(\vec{S}\cdot\tensor{e}\cdot\vec{E})dV$, where $\vec{S}$, $\tensor{e}$ and $\vec{E}$ are the strain field, piezoelectric coupling matrix and electric field. Unlike the capacitive force used in many electro-mechanical systems (see e.g. Refs. \cite{Simmonds_Nature_2011_GroundState, Simmonds_Nature_2011_StrongCoupling}), the piezoelectric force depends on the electric field linearly instead of quadratically \cite{Cleland_NatPhys_2013}. The whole system can be described by the Hamiltonian
\begin{equation}\label{Eq:Hamiltonian}
\begin{aligned}
\hat{H} =& \hbar\Omega_o\hat{a}^\dagger\hat{a} + \hbar\Omega_m\hat{b}^\dagger\hat{b}
+\hbar g_{om}\hat{a}^\dagger\hat{a}(\hat{b}^\dagger+\hat{b}) \\
&+\hbar g_{em}(\hat{c}^\dagger+\hat{c})(\hat{b}^\dagger+\hat{b}) \\
&+i\hbar\sqrt{\kappa_e}(\hat{a}^\dagger\hat{s}_{c,in}e^{-i\Omega_c t} - \hat{a}\hat{s}_{c,in}^\dagger e^{i\Omega_c t}) \\
&+i\hbar\sqrt{\kappa_e}(\hat{a}^\dagger\hat{s}_{p,in}e^{-i\Omega_p t} - \hat{a}\hat{s}_{p,in}^\dagger e^{i\Omega_p t}) ~,
\end{aligned}
\end{equation}
where $\hat{a}$, $\hat{b}$, and $\hat{c}$ are the annihilation operators of the optical, mechanical, and microwave modes. The third and the fourth term describe the opto-mechanical and electro-mechanical coupling characterized by the coupling rates $g_{om}$ and $g_{em}$. The last two terms represent the two optical inputs used in the experiment. $\hat{s}_{c,in}$ ($\hat{s}_{p,in}$) denotes the optical field of the ``control'' (``probe'') light at frequency $\Omega_c$ ($\Omega_p$). Detunings with respect to the optical resonant frequency are defined as $\Delta_c=\Omega_c-\Omega_o$ and $\Delta_p=\Omega_p-\Omega_o$. The output field can be obtained by $\hat{s}_{out}=\hat{s}_{in}-\sqrt{\kappa_e}\hat{a}$.

We study the dynamics of the system using the measurement setup shown in Fig. \ref{fig:device_setup} (d). Light from a tunable laser diode (TLD), used as the control light, is sent to the device through an electro-optical modulator (EOM), which creates sidebands to act as the probe light. A fiber polarization controller (FPC) is used to adjust the laser polarization. In the electrical path, the microwave signal with adjustable amplitude and phase is sent to the device through a transmission line. For the study of higher-harmonic interference, a frequency divider with dividing factor $N$ is applied to the microwave signal, which causes the microwave frequency $\Omega_e$ to be $N$ times less than the control-probe offset frequency, i.e., $\Omega_e=|\Omega_c-\Omega_p|/N$. For now we first consider $N=1$. The optical signal coming out from the device is amplified by an erbium-doped fiber preamplifier (EDFA), filtered by an optical bandpass filter (OBF) and then collected by a photodetector (PD). A wavelength meter is used to calibrate the laser wavelength and a PID feedback control is used to stabilize the laser intensity.

\begin{figure}[t]
\centering
\includegraphics[width=12cm]{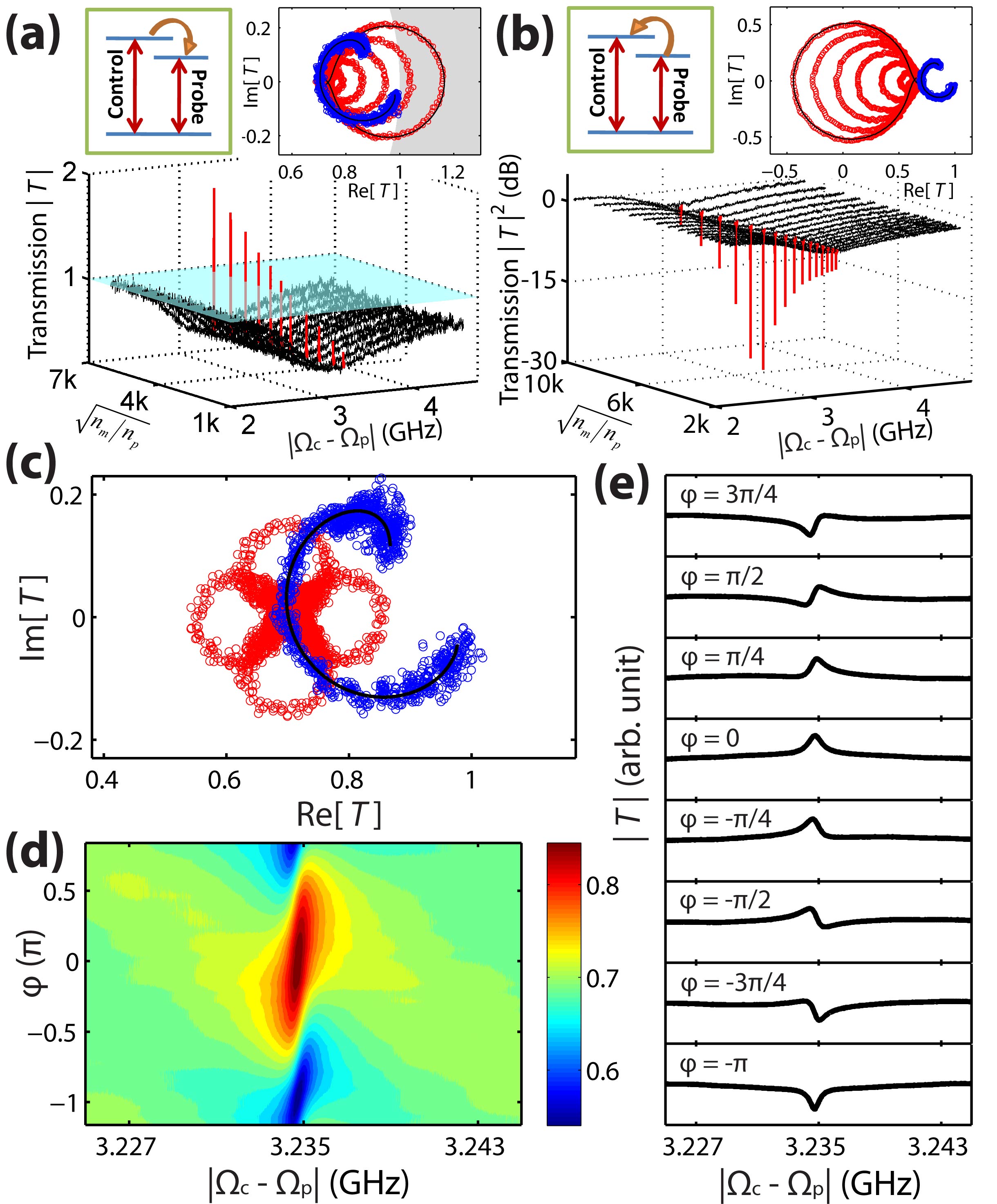}
\caption{(a) \& (b) Probe transmission plotted against $|\Omega_c-\Omega_p|$ as $\sqrt{n_m/n_p}$ 
is varied. The left insets show the energy level diagrams and the right insets plot the transmission coefficient in complex plane (c) $T$ plotted in complex plane as $\varphi$ is varied (d) Contour plot of $|T|$ against $|\Omega_c-\Omega_p|$ and $\varphi$ (e) cross-sectional plot of (d) at various $\varphi$. }
\label{fig:EOMIT}
\end{figure}

We blue-detune the control light with respect to the optical resonance on the shorter wavelength side by a frequency of the third mechanical mode. This mode has a frequency $\Omega_m>\kappa$ and so satisfies the resolved-sideband condition. When the probe light is swept across the optical resonance, the presence of the control light and the microwave actuation opens a transparency window in the originally absorptive region. Fig. \ref{fig:EOMIT} (a) and (b) show the transmission spectra of the probe as the microwave amplitude is varied. The transparency window is highlighted in red color. By adjusting the phase of the microwave mode the interaction can lead to either amplification (Fig. \ref{fig:EOMIT} (a)) or absorption (Fig. \ref{fig:EOMIT} (b)) of the probe light. The processes are analog to the stimulated emission and absorption, as illustrated by the energy-level diagram in the upper-left insets.

The transmission coefficient of the probe light can be derived from the Hamiltonian in Eq. \eqref{Eq:Hamiltonian} using input-output formalism. If the force exerting on the mechanical mode is assumed to be dominated by the piezoelectric force, it can be shown that the probe transmission coefficient $T=s_{p,out}/s_{p,in}$ is given by
\begin{equation}\label{Eq:EIT}
\begin{aligned}
T=&\frac{\kappa/2-\kappa_e-i\Delta_p}{\kappa/2-i\Delta_p} \\
&+\frac{\kappa_e g_{om} g_{em}|c|e^{i\varphi}} {(\kappa/2-i\Delta_p)(\kappa/2-i\Delta_c)(\gamma/2-i\Delta_p)} \frac{s_{c,in}}{s_{p,in}} ~,
\end{aligned}
\end{equation}
where $|c|$ and $\varphi$ are the amplitude and phase of the microwave mode. The first term is the transmission coefficient of the probe in the absence of the control light or the microwave input. Its absolute-squared value gives the typical inverted Lorentzian shape representing the absorption due to the optical resonance. The second term is the interference term which gives rise to the transparency window. When these two terms interfere constructively, increasing the microwave amplitude or the control-to-probe ratio will raise the magnitude of the second term, which will compensate the loss described in the first term, and even amplify the probe signal with an overall gain higher than 1. Fig. \ref{fig:EOMIT} (a) plots the probe transmission as a function of $\sqrt{n_m/n_p}$, where $n_m=|2g_{em}c/\gamma|^2$ and $n_p=|2\sqrt{\kappa_e}s_{p,in}/\kappa|^2$ are the on-resonance phonon and probe photon number. A semi-transparent plane is used to show the level of unity transmission. The upper-right inset plots the transmission coefficient in complex plane, where the blue data points show the overall optical resonance while the red data points show the zoom-in of the transparency window. Since the wide-span data (blue) from different data-sets fall onto the same place, only one of the data-sets is plotted for clarity. The black solid lines show the fitting by Eq. \eqref{Eq:EIT}, taking into account the phase-delay and attenuation background in the measurement. The area with magnitude larger than 1 is shaded in light grey color, indicating the amplification of the probe light. When the phase of the microwave mode $\varphi$ is shifted by $\pi$, the two terms in Eq. \eqref{Eq:EIT} interfere destructively, which causes the originally absorptive region to become even more absorptive. As shown in Fig. \ref{fig:EOMIT} (b), absorption extinction down to 30 dB is achieved while the original cavity absorption extinction is about 3 dB. In this case, the two resonance circles align in opposite direction, as shown in the complex plot in the upper-right inset. At large $\sqrt{n_m/n_p}$ the resonance circle goes beyond the origin causing the group-delay of the probe light to change from negative (advanced light) to positive (slowed light). Slowed light with delay of 0.76 $\mathrm{\mu s}$ and transmission of 20\% is achieved in room temperature.

Besides setting the phase of the microwave mode $\varphi$ to $0$ or $\pi$, $\varphi$ can be tuned continuously from $0$ to $2\pi$ causing the two resonance circles to rotate with respect to each other, as shown in Fig. \ref{fig:EOMIT} (c). As a result, the probe transmission becomes neither a peak nor a dip but a more general asymmetric Fano shape. A contour plot of the probe transmission as a function of $|\Omega_c-\Omega_p|$ and $\varphi$ is shown in Fig. \ref{fig:EOMIT} (d), and its cross-sectional plots at different $\varphi$ are shown in Fig. \ref{fig:EOMIT} (e). With the control of the microwave phase, we are able to tune the Fano resonance through the whole phase-space. Asymmetric Fano resonance is widely observed in many branches of physics (See e.g. Refs. \cite{Chong_NatMater_2010, Kivshar_RMP_2010_FanoNano}), and is also studied recently in the context of optomechanical systems \cite{Agarwal_PRA_2013_OMFano}. It is a manifestation of interference between a continuum and a discrete excitation modes \cite{Fano_PR_1961}. Here, the broader optical resonance takes the role of the continuum mode. 

Compared to the traditional OMIT where the mechanical actuation is from the optical force, here the actuation is provided by the piezoelectric force of the microwave mode \cite{Cleland_NatPhys_2013}. Because of the stronger electrical drive, the transparency phenomenon can be observed with less stringent condition. From Eq. \eqref{Eq:EIT}, it can be shown that the condition for complete transparency is given by $\sqrt{n_m/n_p}(2G/\kappa)\ge 1$, where $G=\sqrt{n_c}g_{om}$ is the modified optomechanical coupling rate. For comparison, condition for the traditional OMIT is $4G^2/\kappa\gamma\gtrsim1$. Here, the effect can be enhanced by increasing the phonon number with coherent microwave drive.

\begin{figure}[t]
\centering
\includegraphics[width=11cm]{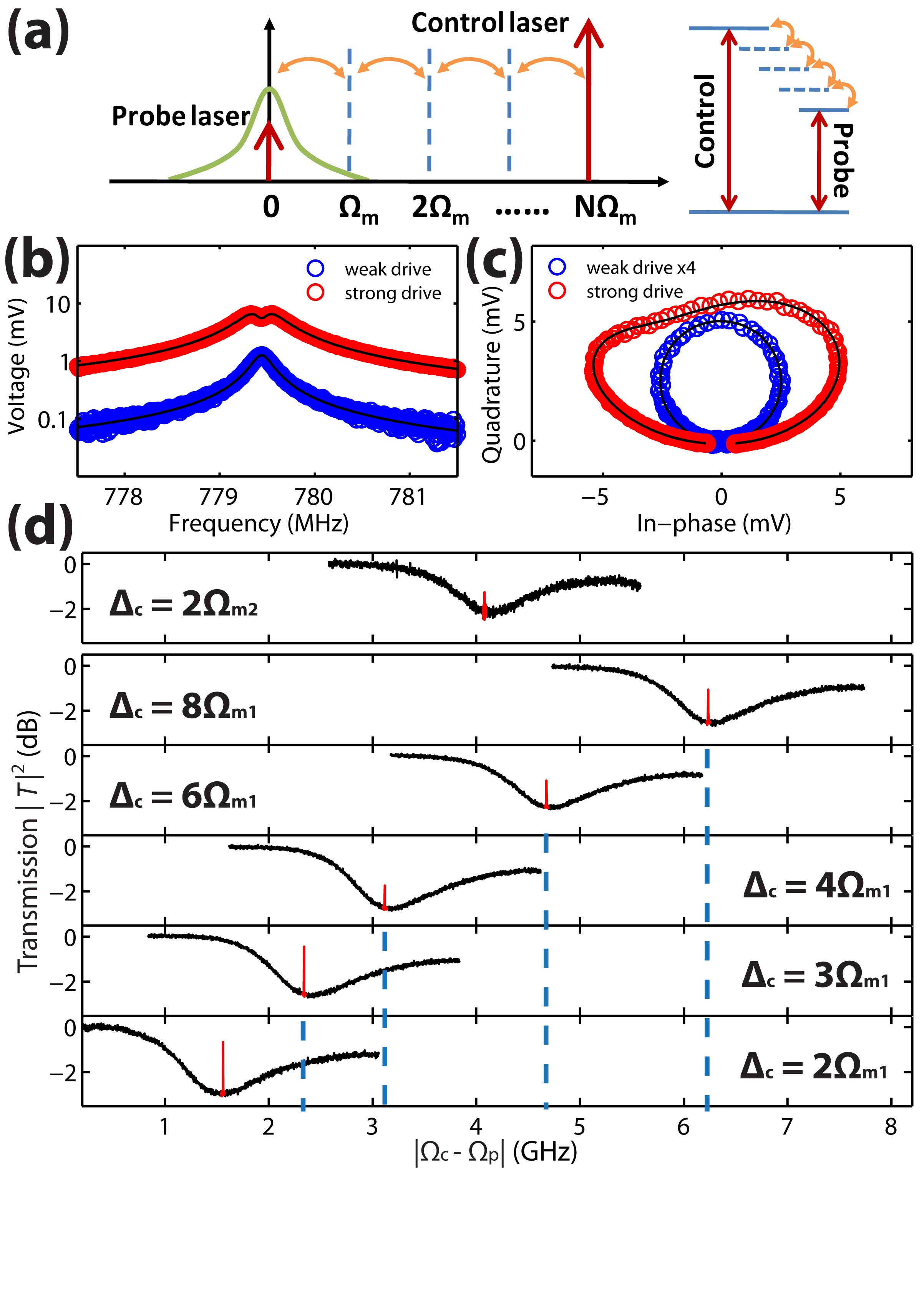}
\caption{ (a) Schematic illustrating the interaction between the control and probe light through multi-phonon scattering. (b) Driven response under weak (blue) and strong (red) microwave drive plotted against frequency. (c) Same data in (b) plotted in complex plane. Data with weak drive (blue) is magnified by 4 times. (d) Transmission spectra of the probe when $\Delta_c$ is at integral multiple of the mechanical resonance frequency.}
\label{fig:harmonics_EOMIT}
\end{figure}

The strong piezoelectric interaction between the microwave and the mechanical mode also facilitates the study of nonlinear dynamics in the optomechanical system. By performing the polaron transformation $\hat{H}^\prime=e^{\hat{S}}\hat{H}e^{-\hat{S}}$ where $\hat{S}=\frac{1}{\Omega_m} [g_{om}\hat{a}^\dagger\hat{a} +g_{em}(\hat{c}^\dagger+\hat{c})] (\hat{b}^\dagger-\hat{b})$ \cite{Girvin_PRL_107_2011, Rabl_PRL_107_2011}, the Hamiltonian in Eq. \eqref{Eq:Hamiltonian} becomes
\begin{equation}\label{Eq:Hamiltonian_transformed}
\begin{aligned}
\hat{H}^\prime =& \hbar\Omega_o\hat{a}^\dagger\hat{a} + \hbar\Omega_m\hat{b}^\dagger\hat{b}
-\frac{\hbar}{\Omega_m}[g_{om}\hat{a}^\dagger\hat{a} +g_{em}(\hat{c}^\dagger+\hat{c})]^2 \\
&+i\hbar\sqrt{\kappa_e} \hat{a}^\dagger\hat{s}_{c,in}e^{-i\Omega_c t} e^{\frac{g_{om}}{\Omega_m}(\hat{b}^\dagger-\hat{b})} + h.c. \\
&+i\hbar\sqrt{\kappa_e} \hat{a}^\dagger\hat{s}_{p,in}e^{-i\Omega_p t} e^{\frac{g_{om}}{\Omega_m}(\hat{b}^\dagger-\hat{b})} + h.c. ~.
\end{aligned}
\end{equation}
When the control light is blue-detuned at $\Delta_c=N\Omega_m$, using rotating wave approximation we can obtain the higher order interaction term proportional to $(\frac{g_{om}}{\Omega_m})^N\hat{a}^\dagger\hat{b}^{\dagger N}+h.c.$, which describes the multi-phonon process as illustrated in the schematic in Fig. \ref{fig:harmonics_EOMIT} (a). This nonlinear interaction term in the Hamiltonian is the key for generating nonclassical states and it leads to high harmonic OMIT in optomechanical systems \cite{Girvin_PRL_111_2013, Clerk_PRL_111_2013}. Such high-order interaction may also lead to coherent multi-boson generation, which can be used to dynamically stabilize the protected cat-qubit encoding \cite{Devoret_arxiv_13122017_2013}. However, the prefactor $(g_{om}/\Omega_m)^N$ is rapidly decaying with $N$ unless the system is close to the single-photon strong-coupling regime with $g_{om}>\Omega_m$ \cite{Girvin_PRL_107_2011}, which is out of reach with the current state-of-the-art. A detail analysis for the case $N=2$ shows that the effect is observable when $g_{om}$ is within a small fraction of $\Omega_m$ and $\kappa$ \cite{Girvin_PRL_111_2013, Clerk_PRL_111_2013}. However higher order interaction with $N>2$ is still inaccessible.

Nevertheless, by utilizing the strong microwave actuation to drive the mechanical resonator into coherent motion, we are able to observe similar nonlinear effect. When the mechanical resonator is in coherent oscillation, we can take $i(\hat{b}^\dagger-\hat{b})=2|b|\sin(\Omega_m t+\theta)$. The translation operator $\exp[{\frac{g_{om}}{\Omega_m} (\hat{b}^\dagger-\hat{b})}]$ in Eq. \eqref{Eq:Hamiltonian_transformed} can then be expressed in terms of Bessel functions using Jacobi-Anger expansion. As a result, at $\Delta_c=N\Omega_m$ and under rotating wave approximation, the $N$-th order term has a dependence of $J_N(\tilde{x})$, which is the Bessel function of the first kind with index $N$ and the argument is given by $\tilde{x}=2|b|g_{om}/\Omega_m$. When $\tilde{x}\ll N$, the Bessel function can be approximated as $J_N(\tilde{x})\approx(\tilde{x}/2)^N/N!$, and so it recovers the $(g_{om}/\Omega_m)^N$ scaling. This approximation breaks down when $\tilde{x}\gtrsim N$. Therefore by driving the mechanical resonator to large coherent amplitude $|b|\approx N\Omega_m/2g_{om}$, the high order nonlinear interaction becomes observable. 

A signature of this large-amplitude-induced nonlinearity can be observed from the mechanical driven response. When only one laser is used to probe the mechanical motion, it can be shown from the transformed Hamiltonian in Eq. \ref{Eq:Hamiltonian_transformed} (or using the approach adopted in Refs. \cite{Girvin_PRL_2006_OMLargeAmp,Tang_PRA_2012_OMLargeAmp}) that the detected driven signal is $\propto(\sum_n J_n(\tilde{x}) J_{n+1}(\tilde{x})/\tilde{x}K_n^\ast K_{n+1})|b|e^{i\theta}$, where $K_n=1-i(\Delta+n\Omega)2/\kappa)$. When $\tilde{x} \ll 1$, the summation term becomes independent of $\tilde{x}$ and so the detected signal is linear to the displacement $|b|e^{i\theta}$. When $\tilde{x}>1$, the driven response becomes amplitude dependent. Fig. \ref{fig:harmonics_EOMIT} (b) plots the driven response of the 1st mechanical mode under weak (blue) and strong (red) drive. Fig. \ref{fig:harmonics_EOMIT} (c) plots the same data in complex plane with the weak-drive data magnified by 4 times. A deviation from the Lorentzian shape (or circular trajectory in complex plane) under strong drive can be clearly observed. The black solid lines are the fitted curves using the full expression, with $g_{om}/\Omega$ and $\Delta/\Omega$ as the fitting parameters.  From the fitting, it can be deduced that $\tilde{x}=1.71$ is reached in the strong-drive data.

To demonstrate optical transparency at higher harmonics using this nonlinear effect, we insert a frequency divider to the microwave path (See Fig. \ref{fig:device_setup} (d)) so that the frequency offset between the probe and control $|\Omega_p-\Omega_c|$ is $N$ times the microwave frequency. The detuning of the control laser is fixed at $\Delta_c=N\Omega_m$. Using the Bessel function expansion as described above, it can be shown that the probe transmission coefficient is given by
\begin{equation}\label{Eq:HHEIT}
\begin{aligned}
T=\frac{\kappa/2-J_0^2(\tilde{x})\kappa_e-i\Delta_p}{\kappa/2-i\Delta_p}
+\frac{\kappa_e J_0(\tilde{x})J_N(\tilde{x}) e^{iN\varphi}}{\kappa/2-i\Delta_p} \frac{s_{c,in}}{s_{p,in}} ~.
\end{aligned}
\end{equation}
The first term is the probe transmission in the absence of the control light. The interference caused by the second term due to the multi-phonon process opens a transparency window at $N$-th harmonic frequency away from the control light. In this measurement we use the first and the second mechanical modes which have lower spring constants and so can be driven into larger amplitude. Fig. \ref{fig:harmonics_EOMIT} (d) shows the high-harmonic optical transparency at various $\Delta_c$ and frequency dividing factor $N$. For the second mechanical mode, transparency window can be observed at the second harmonic frequency ($2f_{m2}=4.08 \mathrm{GHz}$), while for the first mechanical mode, transmission transparency at frequency up to eighth harmonics ($8f_{m1}=6.24 \mathrm{GHz}$) can be observed. This demonstration shows that the coherent microwave drive is a useful tool for studying the nonlinear effect in optomechanical systems. Investigation of the quantum aspect of the effect will be an interesting topic for further study.

In conclusion, we develop a hybrid opto-electro-mechanical system in which the microwave and optical modes are coupled to a common mechanical mode. We demonstrate coherent absorption and amplification, and the more general asymmetric Fano resonance. Using the strong piezoelectric drive, we operate the mechanical mode in large amplitude where high-harmonic transmission transparency through multi-phonon scattering is demonstrated.

We thank Changling Zou and Menno Poot for helpful discussion. We acknowledge funding from DARPA/MTO ORCHID program through a grant from the Air Force Office of Scientific Research (AFOSR). H.X.T. acknowledges support from a Packard Fellowship and a CAREER award from the National Science Foundation. L.J. acknowledges support from the Alfred P. Sloan Foundation, the Packard Foundation, and the DARPA Quiness program.

\bibliography{arxivpreprint_bib}

\end{document}